**Performance gain of displacement receiver with optimized prior probability**

M. Fujiwara[a], S. Izumi[a,b], M. Takeoka[a], and M. Sasaki[a]
a: National Institute of Information and Communications Technology, 4-2-1, Nukui-Kita, Koganei, Tokyo 184-8795, Japan
b: Sophia University, 7-1 Kioicho, Chiyoda-ku, Tokyo 102-8554, Japan

Corresponding author: Mikio Fujiwara
Email: fujiwara@nict.go.jp
Tel: +81-42-327-7552
Fax: +81-42-327-6629

**Abstract:**
We numerically study the performance of the displacement based quantum receiver for the discrimination of weak 3- and 4-phase-shift keyed (PSK) coherent state signals. We show that due to the nontrivial asymmetry of the receiver structure, optimization of the prior probability increases the mutual information and achieves sub-shot-noise limit discrimination. Moreover, we estimate the cutoff rate for a 4-PSK signal and confirm that the prior probability optimization shortens the code length for a given decoding error criterion. Such consideration for the asymmetric channel matrix is essential in a study of the compassable quantum receiver.

## 1. Introduction

In a highly attenuated optical communication channel, the quantum noise fundamentally prevents perfect discrimination of optical signals. The conventional theoretical limit of coherent state signals is known as the standard quantum limit (SQL), and it can be reached by an ideal heterodyne or homodyne detection schemes. On the other hand, quantum theory enables us to obtain error rates smaller than the SQL. Here, we refer to the receiver outperforming the SQL as ‘quantum receiver’. Physical implementation of quantum receivers for a binary phase-shift keyed (BPSK) signal was first proposed in the 1970s [1, 2]. These schemes consist of photon counting and phase space displacement operation and recent progress on these technologies enabled experimental demonstrations [3-6]. In particular, the near-optimal quantum receiver called the optimal displacement receiver, which is a generalized version of [1], was proposed recently [7, 8] and successfully demonstrated the sub-SQL bit error rate discrimination of the BPSK signals without compensation of any technical imperfections [6].

The quantum receiver for more than two signals has also been investigated. Bondurant proposed a receiver composed of a single photon detector and feedback circuit for discrimination of quadrature phase shift keyed (QPSK, 4-PSK) signals comprising of four equally phase separated states [9]. Mueller et al. proposed and experimentally demonstrated a discrimination of QPSK signals with a hybrid receiver that executes sequential detection of a heterodyne detector and a displacement receiver [10]. Recently a multiple-stage displacement receiver based on the feedforward principle for 3-PSK and 4-PSK signals has been proposed [11, 12]. In the discrimination process, the amount of displacement is determined from the results of the previous stages. They showed that a feedforward process of several steps could improve the bit error rate so as to overcome the SQL when the received signal contains only a few photons.

These previous works on quantum receivers deal with the average bit error rate (BER) of signal discrimination. However, in the context of practical communications, not only the average BER but also the decoding error probability in discriminating code words (block sequences consisting of the signals), and eventually the transmission rate of asymptotically error-free communication should be important concerns. Such a transmission rate is measured based on information theoretic quantities, such as the mutual information and cutoff rate which are defined for a given channel matrix. Usually the mutual information is maximized in terms of the prior probability distribution of the signal set to evaluate the channel capacity. In quantum communication, on the other hand, a physical model of transmission channel is initially given, i.e., before the channel matrix is specified. In designing quantum communication system, the measurement can be optimized for the output signals from the transmission channel. The channel matrix is determined after an appropriate measurement is specified. Thus, the range of optimization is widened so as to vary detection strategy as well. Thus, an important concern has been finding and implementing optimal quantum receivers and quantum decoders.

Unfortunately, however, though there are few theoretical proposals of the optimal schemes, its experimental implementing is usually not an easy task. Currently available quantum receivers in laboratories are more or less the displacement receiver, which is suboptimal but relatively easy to implement. This receiver often produces an asymmetric output probability distribution even for a symmetric signal set with an equiprobable prior probability distribution, and hence results in an asymmetric channel matrix. Therefore, a practical concern in the application of quantum receiver to optical communication systems is the maximization of the mutual information and cutoff rate in terms of the prior probability distribution for a given asymmetric channel matrix based on the displacement receiver.

In this paper, we investigate this problem numerically in the cases of 3- and 4- PSK coherent state signals, especially in regions of weak coherent states where the quantum noise essentially limits the communication performances. Assuming PSK weak coherent state signals and the displacement receiver, we evaluate the channel capacity defined as the mutual information maximized over the prior probability distribution. Finally, we estimate the cutoff rate and decoding error. We show that the optimization of the prior probability is essential for better design of the communication system based on the displacement receiver.

## 2. Displacement receiver

The physical set up of our displacement receiver is the same as that in Ref. [12]. We briefly explain its signal detection strategy.

The M-ary PSK coherent states $|\alpha_m>$, m=0, 1, 2,..., M-1, are defined as

$$|\alpha_m\rangle = |\alpha u^m\rangle, u = \exp\left(\frac{2\pi i}{M}\right), \tag{1}$$

where without loss of generality, α is chosen to be a real number. Alternatively the states can be represented as

$$|\alpha_m\rangle = \hat{V}^m|\alpha_0\rangle, \hat{V} = \exp\left(\frac{2\pi i}{M}\hat{n}\right), \tag{2}$$

where $\hat{n}$ represents the photon number operator. The displacement receiver consists of beam splitters, displacement and on-off detectors. Displacement operation shifts the amplitude of coherent state as $\hat{D}(\alpha)|\beta\rangle = |\beta + \alpha\rangle$. The displacement operation is achieved by combining the signal and a local oscillator via a highly transmissive beam splitter. The on-off detector is a photon detection device observing zero or nonzero photons. The on-off detector is described by a set of operators:

$$\hat{\Pi}_{off} = \exp(-\gamma)\sum_{n=0}^{\infty}(1-\eta)^n|n\rangle\langle n|$$

$$\hat{\Pi}_{on} = 1 - \hat{\Pi}_{off}, \quad (3)$$

where $\Upsilon$ is the dark count probability and $\eta$ is the detection efficiency. The probability of finding an off signal when detecting $|\alpha_m>$ is given by

$$P = \langle \alpha_m | \hat{\Pi}_{off} | \alpha_m \rangle = \exp(-\gamma - \eta\alpha^2). \quad (4)$$

The displacement receiver with feedforward consists of the cascade of displacement and photon detection. The input signal is divided into the number of feedforward steps via a beam splitter. In each step, the split signal is displaced by $\hat{D}\left(-\sqrt{1-r}\alpha_i\right)$, where r and $\alpha_i$ are the reflectance of the beam splitter and the prepared coherent state of the branch. It is then detected by the on-off detector. If the split signal matches the prepared state and the on-off detector is perfect, the firing probability of the on-off detector becomes equal to zero. At each feedforward step, the displacement value is decided in accordance with the outcome of the photon detection in the previous step. The state discrimination is accomplished with the combination of detected signals of all branches. The three-step feedforward tree for 4-PSK signal is shown in Fig. 1.

In our calculation, we assume a three-step feedforward discrimination and considering technical viability. The channel matrices are given in the Appendix. We assume a unit quantum efficiency and the dark count probability of 10^-8 for the on-off photon detector and the 100% visibility of the interferometer. In the discrimination of a 4-PSK signal, the order of displacing the signal states affects the performance of the receiver. We therefore carefully choose the order of the displacement to maximize the performance (see Fig. 1). We first displace the state of symbol ‘0’. If the detector clicks in the first step, the displacement value of the second step is adjusted to discriminate the diagonal symbol of ‘2’. On the other hand, in the case of 3-PSK signals, the performance is independent of the order of displacement since the distances between signals are all equal. If the output matches the expected input, that is, the one which is displaced in the first step, then no displacement is made in the second step, in terms of the likelihood of the input being higher. The same evaluation is repeated twice (total 3 times). Thus, the most reliable level of discrimination is achieved.

Given a channel matrix [P(y|x)] between input signals {x} and output ones {y}, and prior probabilities [P(x)], the mutual information is given by

$$I(X:Y) = \sum_x P(x) \sum_y P(y|x) \log_2 \left[\frac{P(y|x)}{\sum_{x'} P(x')P(y|x')}\right]. \quad (5)$$

In the next section, we discuss the maximization of the mutual information of the displacement receiver in terms of P(x).

## 3. Optimization of the prior probability

We use *Mathematica* to compute the optimal prior probability. The mutual information as a function of the received photon number for the displacement receiver with the prior probability optimized and equal for (a) 3-PSK and (b) 4-PSK signals is shown in Fig. 2. The mutual information obtained by a heterodyne detector is also plotted. The optimization of the prior probability results in an increase in the mutual information at the received power region of a few photons. In the high-input power region, the optimized mutual information comes close to that for an equal prior probability. In the case of a 4-PSK signal discrimination, the improvement in the mutual information is more apparent, i.e., the gain can be recognized with at least four photons.

The bit error rates of the displacement receiver with the prior probability optimized such that the mutual information is maximized are shown in Figs. 3(a) and (b) for 3-PSK and 4-PSK discrimination. Also, plotted are the displacement receiver, SQL attained by heterodyne receiver, and the Helstrom limit [13] for equal prior probabilities. Sudden drops of the error rate at a photon number very near zero in both figures reflect the instability of the numerical optimization.

If the prior probability is set to be equal to the 4-PSK signal, the error rate of the displacement receiver is worse than that of a heterodyne detector when the input photon number falls below two. On the other hand, the optimization of the prior probability enables the displacement receiver to obtain a lower bit error rate than the SQL even when the received power is less than two photons. The curve has a discontinuous shape. It might imply that the optimization of the prior probability is incomplete. However, it is clear that the asymmetric prior probability in our receiver gives a lower bit error rate. The optimized prior probabilities are shown in Fig. 4(a) and (b) for 3-PSK and 4-PSK discrimination, respectively. As expected, one can see that the prior probability of the first evaluated symbol stands out in the small input photon region. For 3-PSK signals, the prior probabilities of the other symbols have similar curves. On the other hand, the prior probabilities of the first two symbols increase in the 4-PSK discrimination process in the heavily attenuated region. The mutual information near the vacuum region in which the input photon number is less than 0.5 photons is maximized by shifting a signal format from 4-PSK to BPSK shown in Fig. 4 (b). The kink in Fig. 3(b) corresponds to points in which prior probabilities of two symbols reach 0 in Fig. 4(b). As shown in Fig. 4(b), some nodes appear in the prior probability curve. This is probably due to the imperfection of optimization when the optimal signal set has changed from the 4-PSK set to the BPSK set. Such a signal format replacement improves the quantity of the mutual information corresponding to the decrease of error rate shown in Fig. 3(b) due to the asymmetric property of the displacement receiver. In other words, using the displacement receiver could lead to a replacement and thus achieve efficient communication through a strongly attenuated and variable transmission channel.

We also estimate the cutoff rate [13] of the displacement receiver. This rate is a multi-faceted parameter of communication that owes its significance in part to its relation to sequential decoding. Originally, the optimization of the prior probability according to the equation (6) is an essential operation to obtain the cutoff rate, Rc.

$$Rc = -\ln\{\min_{\{pi\}} \left[ \sum_{j=1}^{n} \left( \sum_{i=1}^{m} p_i \{P(j|i)\}^{1/2} \right)^2 \right]\} \qquad (6)$$

where $p_i$ is the prior probability, and P(i|j) is the conditional probability that the *j*th symbol is detected at the receiver side when the sender transmits the *i*th symbol. Using Rc enables us to evaluate a practical limitation of the communication performance in the region of medium transmission rates.

The cut-off rates derived by numerically optimizing the prior probability are plotted as a function of average input photon number in Fig. 5(a) and (b). For reference, Rc with the prior probabilities optimized to maximize the mutual information and set to be equal for all the signals are also plotted in the Fig. 5(a). We observe that the former plot almost coincides with the actual cutoff rate. Moreover, the optimized prior probability of the cutoff rate shown in Fig. 5(b) shows a similar qualitative trend with that optimized for the mutual information. However, there is a slight difference between them. This variance reflects the difference of code lengths assumed in the mutual information and the cutoff rate. For a finite code length, the decoding error is upper bounded by $\exp(-NR_c)$, where N is the code length. The upper bound of the decoding error obtained by Rc with the optimal and equal prior probabilities are shown in Fig. 5(c).We assume the 4-PSK signals with an average photon number at the receiver (Fig. 5(b)). Compared with the equal prior probability, the code length needed to achieve the error-free communication (bit error rate less than 10^-9) can be decreased by about 38% after the optimization.

## 4. Summary

We numerically confirm that the mutual information of the displacement receiver with a three-step feed forward process can be increased by optimizing the prior probability for 3- and 4-PSK signals at a few photon levels. Needless to say, a multi-phase shift keyed signal can transmit data efficiently with the same energy at a certain received power level. Preparation of a receiver for multi-phase shift encoded signals would be necessary for efficient communication over the entire received power range. Our results indicate that a displacement receiver for multi-phase shift keyed signals can head off decreases in signal transmission performance by changing the prior probability in heavily attenuated input power regions. This strategy is feasible for realizing an adaptive receiver that has a performance in tune with the conditions of the communication channel. To realize the quantum receiver, using a detection strategy based on the displacement receiver with the asymmetric structure under the existing condition is both realistic and efficient.

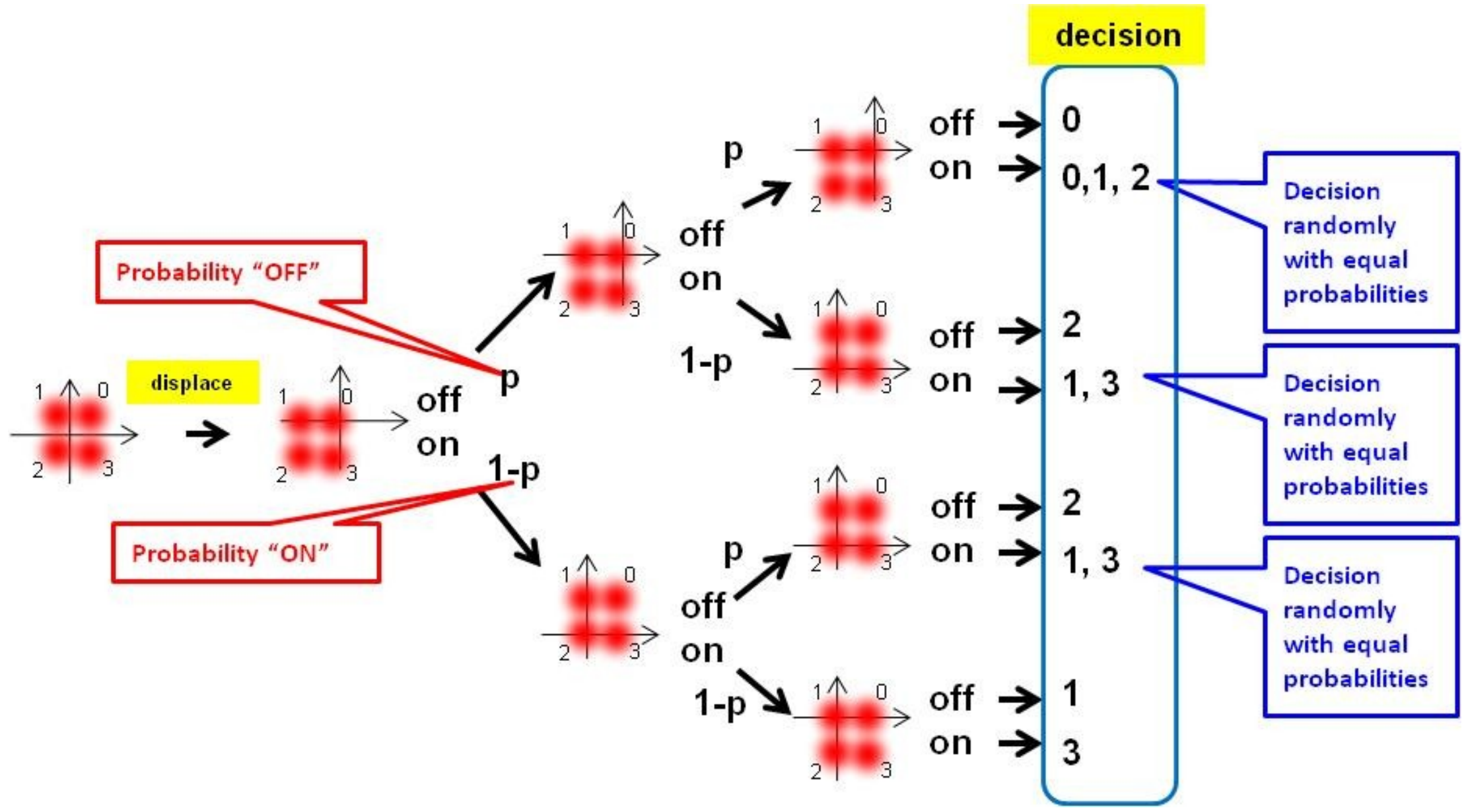

Fig. 1. The displacement receiver with a three-step feedforward tree for 4-PSK signal. Red circles show coherent states, and 0 to 3 correspond to modulated numbers.
P is probability of finding off signal of on-off detector.

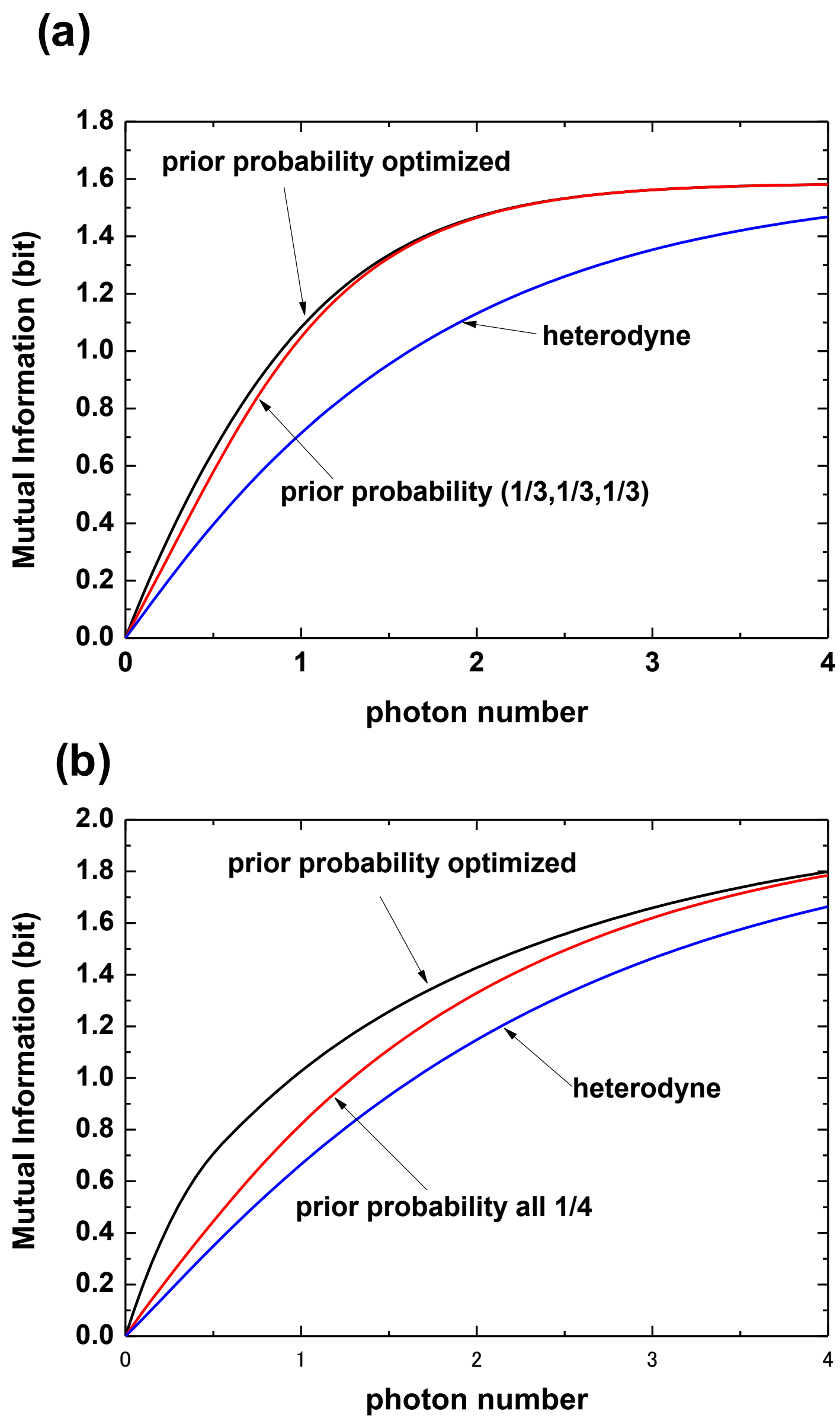

Fig. 2. Mutual information as functions of input photon number for cases of optimized (black curve) and equal prior probability (red curve) for (a) 3-PSK and (b) 4-PSK. Results of heterodyne detection are also plotted (blue curve).

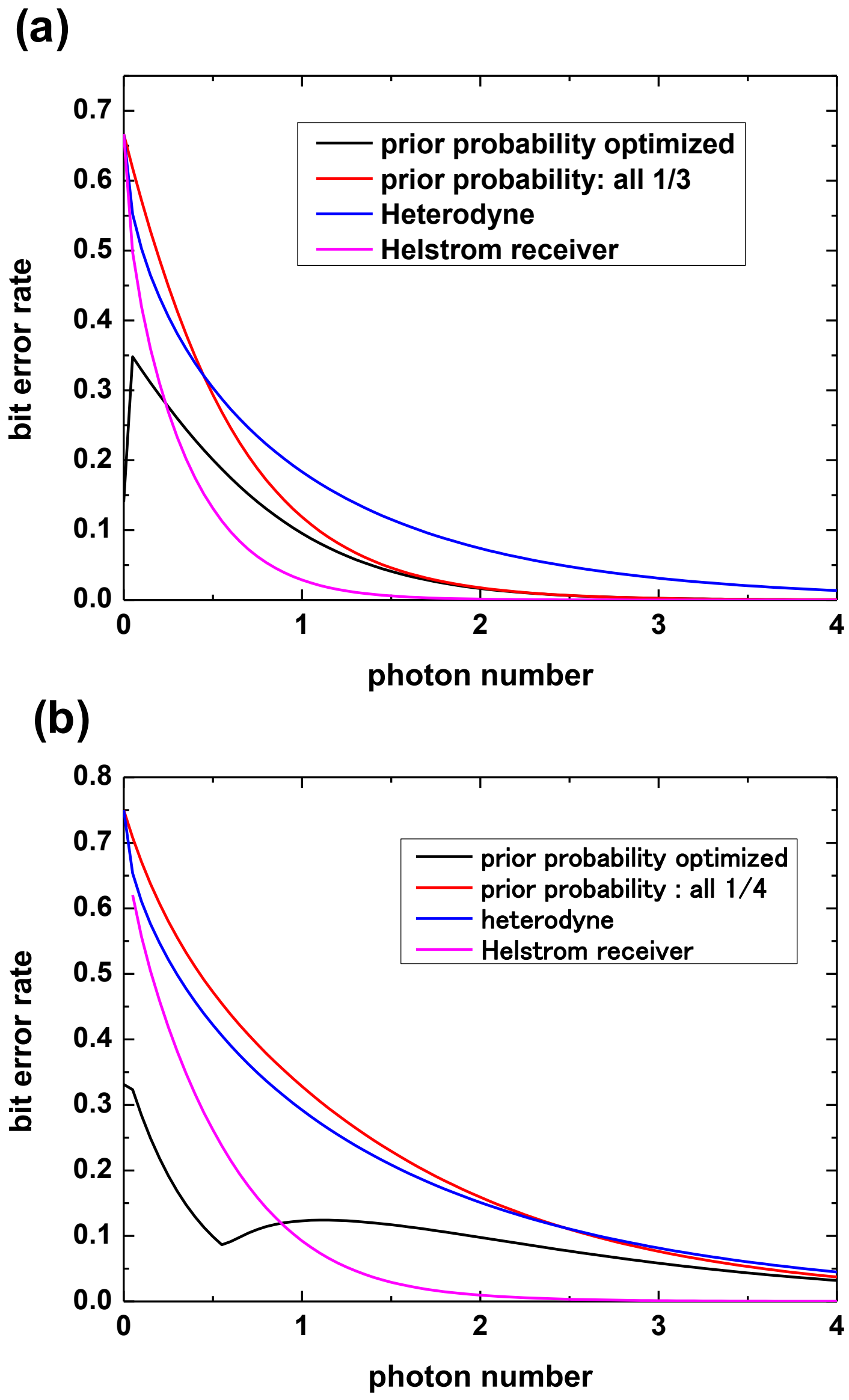

Fig. 3. Bit error rates as functions of input photon number for prior optimized (black curve) and equal probability (red curve), heterodyne (blue curve), and Helstrom limit (crimson curve) for (a) 3-PSK and (b) 4-PSK.

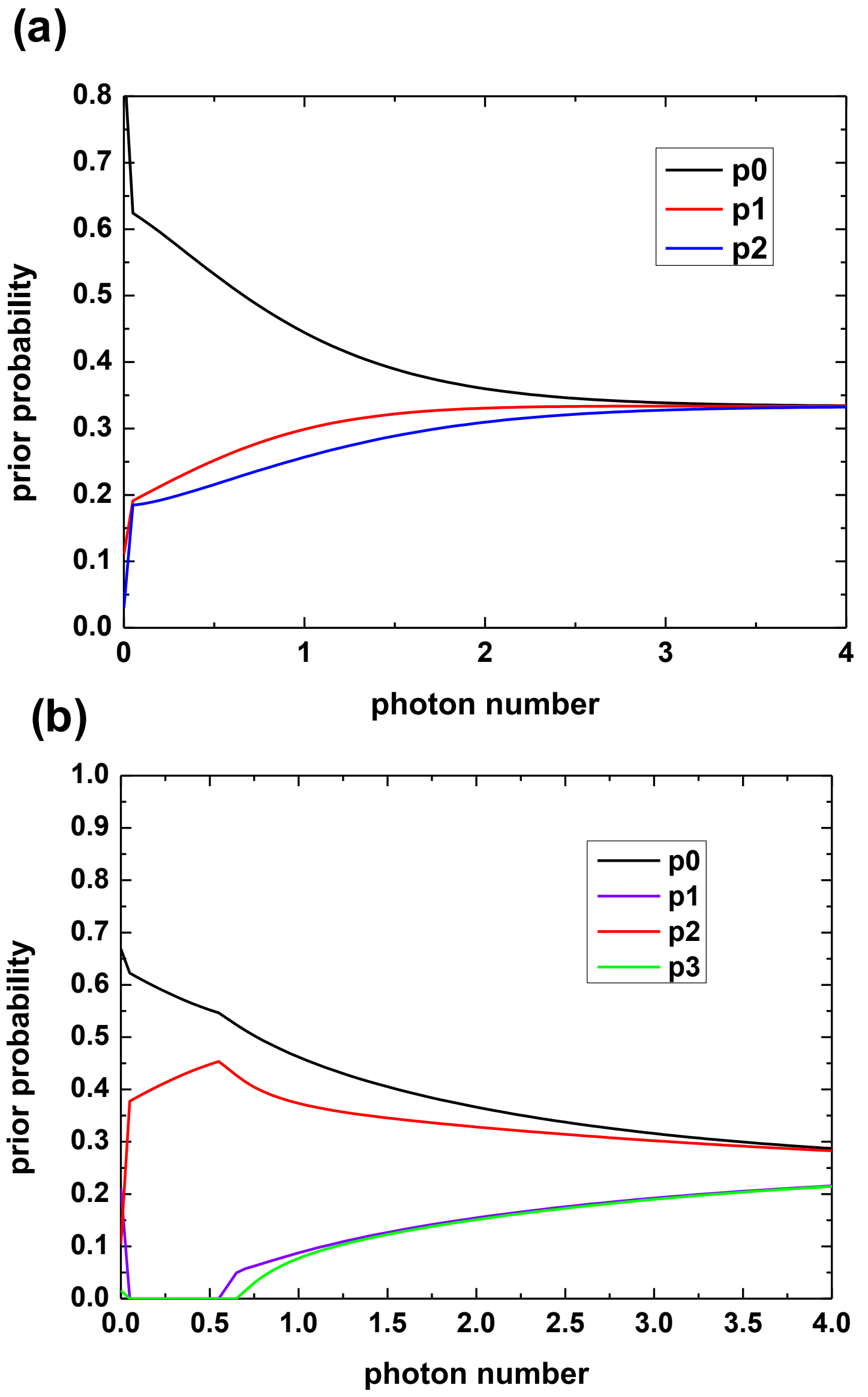

Fig. 4. Prior probability optimized using mutual information for (a) 3-PSK and (b) 4-PSK. Here, “p0”, “p1”, “p2”, and “p3” are prior probabilities of each symbol respectively.

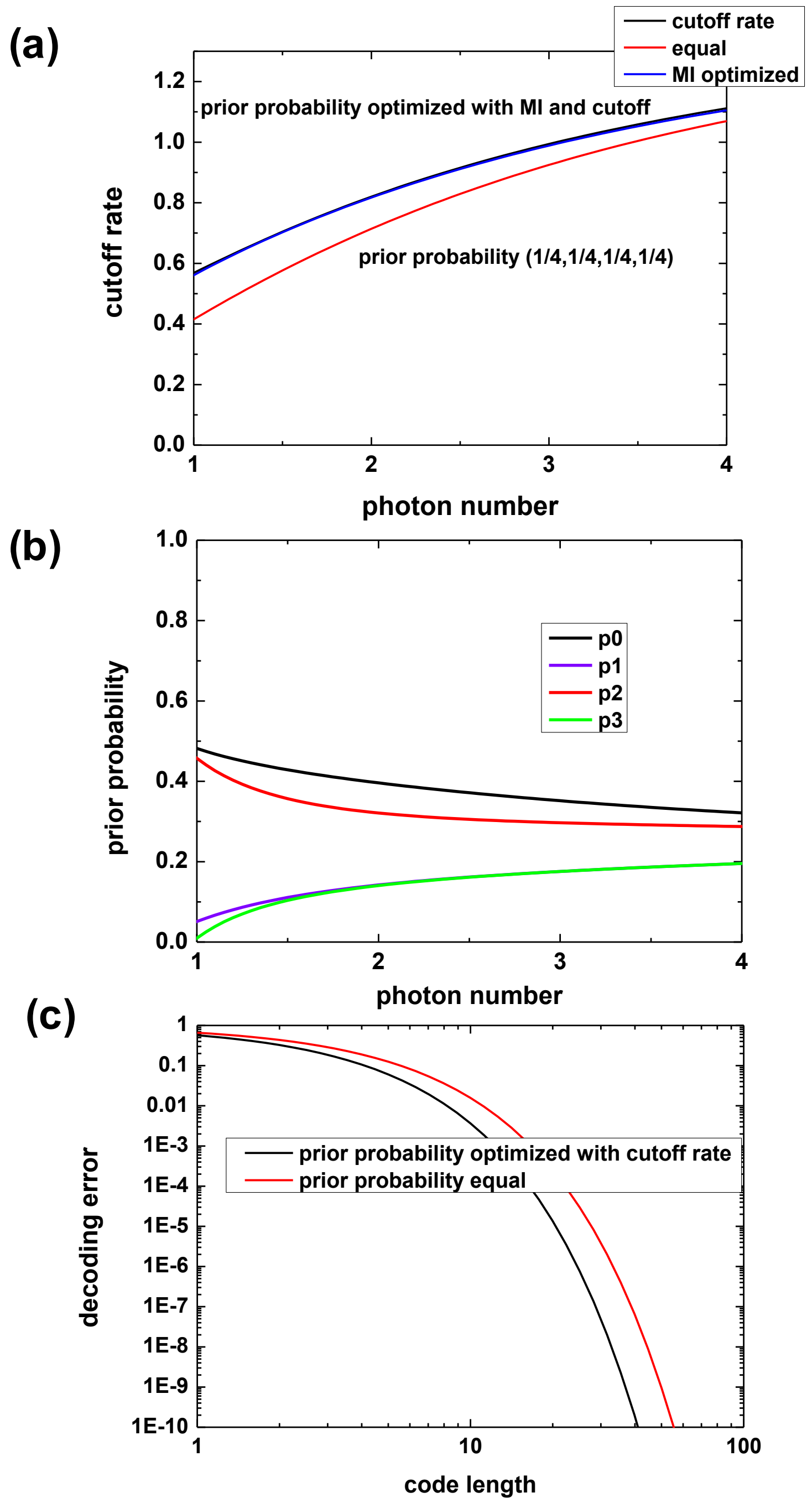

Fig. 5.(a) Cutoff rates of displacement receiver for 4-PSK signal (black curve). For reference, rate with the prior probability optimized by mutual information (blue curve) and equal probability (red curve) are plotted. In (b) prior probability of the cutoff rate as function of input photon number is shown. As reference, rate with equal prior probability is also plotted. (c) Decoding error as function of code length (photon number: 1) with prior probability optimized using cutoff rate (black curve) and equal prior probability (red curve).

APPENDIX: Channel matrix of displacement receiver with 3-step feed forward

In this appendix, we give the channel matrix of the displacement receiver with three step feedforward. In the equation, α, ϒ, and η correspond to the amplitude of the coherent state, the dark count, and detection efficiencies. Also, $r_1$ and $r_2$ are reflection rates of beam splitters in the displacement receiver shown in Fig. A1. We assume $r_1$ and $r_2$ are 2/3 and 1/2 respectively. The conditional probabilities for 4-PSK and 3-PSK signals are given as follows.

Channel matrix for 4-PSK signal

$$P(0|0) = e^{-\gamma} \cdot e^{-\gamma} \cdot e^{-\gamma}$$

$$\begin{aligned} P(1|0) = {} & \tfrac{1}{3} e^{-\gamma} \cdot e^{-\gamma} \cdot (1 - e^{-\gamma}) + \tfrac{1}{2} e^{-\gamma} \cdot (1 - e^{-\gamma}) \cdot \left(1 - e^{-\gamma - 4r_1(1-r_2)\eta\alpha^2}\right) \\ & + \tfrac{1}{2}(1 - e^{-\gamma}) \cdot e^{-\gamma - 4r_1 r_2 \eta\alpha^2} \cdot \left(1 - e^{-\gamma - 4r_1(1-r_2)\eta\alpha^2}\right) \\ & + (1 - e^{-\gamma}) \cdot e^{-\gamma - 4r_1 r_2 \eta\alpha^2} \cdot e^{-\gamma - 2r_1(1-r_2)\eta\alpha^2} \end{aligned}$$

$$\begin{aligned} P(2|0) = {} & \tfrac{1}{3} e^{-\gamma} \cdot e^{-\gamma} \cdot (1 - e^{-\gamma}) + e^{-\gamma} \cdot (1 - e^{-\gamma}) \cdot \left(1 - e^{-\gamma - 4r_1(1-r_2)\eta\alpha^2}\right) \\ & + (1 - e^{-\gamma}) \cdot e^{-\gamma - 4r_1 r_2 \eta\alpha^2} \cdot e^{-\gamma - 4r_1(1-r_2)\eta\alpha^2} \end{aligned}$$

$$\begin{aligned} P(3|0) = {} & \tfrac{1}{3} e^{-\gamma} \cdot e^{-\gamma} \cdot (1 - e^{-\gamma}) + \tfrac{1}{2} e^{-\gamma} \cdot (1 - e^{-\gamma}) \cdot \left(1 - e^{-\gamma - 4r_1(1-r_2)\eta\alpha^2}\right) \\ & + \tfrac{1}{2}(1 - e^{-\gamma}) \cdot e^{-\gamma - 4r_1 r_2 \eta\alpha^2} \cdot \left(1 - e^{-\gamma - 4r_1(1-r_2)\eta\alpha^2}\right) \\ & + (1 - e^{-\gamma}) \cdot \left(e^{-\gamma - 4r_1 r_2 \eta\alpha^2}\right) \cdot \left(1 - e^{-\gamma - 2r_1(1-r_2)\eta\alpha^2}\right) \end{aligned}$$

$$P(0|1) = e^{-\gamma - 2(1-r_1)\eta\alpha^2} \cdot e^{-\gamma - 2r_1 r_2 \eta\alpha^2} \cdot e^{-\gamma - 2r_1(1-r_2)\eta\alpha^2}$$

$$P(1|1) = \frac{1}{3}e^{-\gamma-2(1-r_1)\eta\alpha^2} \cdot e^{-\gamma-2r_1r_2\eta\alpha^2} \cdot \left(1 - e^{-\gamma-2r_1(1-r_2)\eta\alpha^2}\right) + \frac{1}{2}e^{-\gamma-2(1-r_1)\eta\alpha^2} \cdot \left(1 - e^{-\gamma-2r_1r_2\eta\alpha^2}\right) \cdot \left(1 - e^{-\gamma-2r_1(1-r_2)\eta\alpha^2}\right) + \frac{1}{2}\left(1 - e^{-\gamma-2(1-r_1)\eta\alpha^2}\right) \cdot e^{-\gamma-2r_1r_2\eta\alpha^2} \cdot \left(1 - e^{-\gamma-2r_1(1-r_2)\eta\alpha^2}\right) + \left(1 - e^{-\gamma-2(1-r_1)\eta\alpha^2}\right) \cdot \left(1 - e^{-\gamma-2r_1r_2\eta\alpha^2}\right) \cdot e^{-\gamma}$$

$$P(2|1) = \frac{1}{3}e^{-\gamma-2(1-r_1)\eta\alpha^2} \cdot e^{-\gamma-2r_1r_2\eta\alpha^2} \cdot \left(1 - e^{-\gamma-2r_1(1-r_2)\eta\alpha^2}\right) + e^{-\gamma-2(1-r_1)\eta\alpha^2} \cdot \left(1 - e^{-\gamma-2r_1r_2\eta\alpha^2}\right) \cdot e^{-\gamma-2r_1(1-r_2)\eta\alpha^2} + \left(1 - e^{-\gamma-2(1-r_1)\eta\alpha^2}\right) \cdot e^{-\gamma-2r_1r_2\eta\alpha^2} \cdot e^{-\gamma-2r_1(1-r_2)\eta\alpha^2}$$

$$P(3|1) = \frac{1}{3}e^{-\gamma-2(1-r_1)\eta\alpha^2} \cdot e^{-\gamma-2r_1r_2\eta\alpha^2} \cdot \left(1 - e^{-\gamma-2r_1(1-r_2)\eta\alpha^2}\right) + \frac{1}{2}e^{-\gamma-2(1-r_1)\eta\alpha^2} \cdot \left(1 - e^{-\gamma-2r_1r_2\eta\alpha^2}\right) \cdot \left(1 - e^{-\gamma-2r_1(1-r_2)\eta\alpha^2}\right) + \frac{1}{2}\left(1 - e^{-\gamma-2(1-r_1)\eta\alpha^2}\right) \cdot e^{-\gamma-2r_1r_2\eta\alpha^2} \cdot \left(1 - e^{-\gamma-2r_1(1-r_2)\eta\alpha^2}\right) + \left(1 - e^{-\gamma-2(1-r_1)\eta\alpha^2}\right) \cdot \left(1 - e^{-\gamma-2r_1r_2\eta\alpha^2}\right) \cdot (1 - e^{-\gamma})$$

$$P(0|2) = e^{-\gamma-4(1-r_1)\eta\alpha^2} \cdot e^{-\gamma-4r_1r_2\eta\alpha^2} \cdot e^{-\gamma-4r_1(1-r_2)\eta\alpha^2}$$

$$P(1|2) = \frac{1}{3}e^{-\gamma-4(1-r_1)\eta\alpha^2} \cdot e^{-\gamma-4r_1r_2\eta\alpha^2} \cdot (1 - e^{-\gamma-4r_1(1-r_2)\eta\alpha^2}) + \frac{1}{2}e^{-\gamma-4(1-r_1)\eta\alpha^2} \cdot \left(1 - e^{-\gamma-4r_1r_2\eta\alpha^2}\right) \cdot (1 - e^{-\gamma}) + \frac{1}{2}(1 - e^{-\gamma-4(1-r_1)\eta\alpha^2}) \cdot e^{-\gamma} \cdot (1 - e^{-\gamma}) + (1 - e^{-\gamma-4(1-r_1)\eta\alpha^2}) \cdot (1 - e^{-\gamma}) \cdot e^{-\gamma-2r_1(1-r_2)\eta\alpha^2}$$

$$P(2|2) = \frac{1}{3}e^{-\gamma-4(1-r_1)\eta\alpha^2} \cdot e^{-\gamma-4r_1r_2\eta\alpha^2} \cdot (1 - e^{-\gamma-4r_1(1-r_2)\eta\alpha^2}) + e^{-\gamma-4(1-r_1)\eta\alpha^2} \cdot \left(1 - e^{-\gamma-4r_1r_2\eta\alpha^2}\right) \cdot e^{-\gamma} + (1 - e^{-\gamma-4(1-r_1)\eta\alpha^2}) \cdot e^{-\gamma} \cdot e^{-\gamma}$$

$$P(3|2) = \frac{1}{3}e^{-\gamma-4(1-r_1)\eta\alpha^2} \cdot e^{-\gamma-4r_1r_2\eta\alpha^2} \cdot (1 - e^{-\gamma-4r_1(1-r_2)\eta\alpha^2}) + \frac{1}{2}e^{-\gamma-4(1-r_1)\eta\alpha^2}$$
$$\cdot \left(1 - e^{-\gamma-4r_1r_2\eta\alpha^2}\right) \cdot (1 - e^{-\gamma}) + \frac{1}{2}(1 - e^{-\gamma-4(1-r_1)\eta\alpha^2}) \cdot e^{-\gamma}$$
$$\cdot (1 - e^{-\gamma}) + (1 - e^{-\gamma-4(1-r_1)\eta\alpha^2}) \cdot (1 - e^{-\gamma}) \cdot (1 - e^{-\gamma-2r_1(1-r_2)\eta\alpha^2})$$

$$P(0|3) = e^{-\gamma-2(1-r_1)\eta\alpha^2} \cdot e^{-\gamma-2r_1r_2\eta\alpha^2} \cdot e^{-\gamma-2r_1(1-r_2)\eta\alpha^2}$$

$$P(1|3) = \frac{1}{3}e^{-\gamma-2(1-r_1)\eta\alpha^2} \cdot e^{-\gamma-2r_1r_2\eta\alpha^2} \cdot \left(1 - e^{-\gamma-2r_1(1-r_2)\eta\alpha^2}\right) + \frac{1}{2}e^{-\gamma-2(1-r_1)\eta\alpha^2}$$
$$\cdot \left(1 - e^{-\gamma-2r_1r_2\eta\alpha^2}\right) \cdot \left(1 - e^{-\gamma-2r_1(1-r_2)\eta\alpha^2}\right)$$
$$+ \frac{1}{2}\left(1 - e^{-\gamma-2(1-r_1)\eta\alpha^2}\right) \cdot e^{-\gamma-2r_1r_2\eta\alpha^2} \cdot \left(1 - e^{-\gamma-2r_1(1-r_2)\eta\alpha^2}\right)$$
$$+ \left(1 - e^{-\gamma-2(1-r_1)\eta\alpha^2}\right) \cdot (1 - e^{-\gamma-2r_1r_2\eta\alpha^2}) \cdot e^{-\gamma-4r_1(1-r_2)\eta\alpha^2}$$

$$P(2|3) = \frac{1}{3}e^{-\gamma-2(1-r_1)\eta\alpha^2} \cdot e^{-\gamma-2r_1r_2\eta\alpha^2} \cdot \left(1 - e^{-\gamma-2r_1(1-r_2)\eta\alpha^2}\right) + e^{-\gamma-2(1-r_1)\eta\alpha^2}$$
$$\cdot \left(1 - e^{-\gamma-2r_1r_2\eta\alpha^2}\right) \cdot \left(1 - e^{-\gamma-2r_1(1-r_2)\eta\alpha^2}\right)$$
$$+ \left(1 - e^{-\gamma-2(1-r_1)\eta\alpha^2}\right) \cdot e^{-\gamma-2r_1r_2\eta\alpha^2} \cdot e^{-\gamma-2r_1(1-r_2)\eta\alpha^2}$$

$$P(3|3) = \frac{1}{3}e^{-\gamma-2(1-r_1)\eta\alpha^2} \cdot e^{-\gamma-2r_1r_2\eta\alpha^2} \cdot \left(1 - e^{-\gamma-2r_1(1-r_2)\eta\alpha^2}\right) + \frac{1}{2}e^{-\gamma-2(1-r_1)\eta\alpha^2}$$
$$\cdot \left(1 - e^{-\gamma-2r_1r_2\eta\alpha^2}\right) \cdot \left(1 - e^{-\gamma-2r_1(1-r_2)\eta\alpha^2}\right)$$
$$+ \frac{1}{2}\left(1 - e^{-\gamma-2(1-r_1)\eta\alpha^2}\right) \cdot e^{-\gamma-2r_1r_2\eta\alpha^2} \cdot \left(1 - e^{-\gamma-2r_1(1-r_2)\eta\alpha^2}\right)$$
$$+ \left(1 - e^{-\gamma-2(1-r_1)\eta\alpha^2}\right) \cdot \left(1 - e^{-\gamma-2r_1r_2\eta\alpha^2}\right) \cdot (1 - e^{-\gamma-4r_1(1-r_2)\eta\alpha^2})$$

(a-1)

Channel matrix for 3-PSK signal

$$P(0|0) = e^{-\gamma} \cdot e^{-\gamma} \cdot e^{-\gamma}$$

$$P(1|0) = \frac{1}{2}e^{-\gamma} \cdot e^{-\gamma} \cdot (1 - e^{-\gamma}) + e^{-\gamma} \cdot (1 - e^{-\gamma}) \cdot e^{-\gamma - 3r_1 r_2 \eta\alpha^2} + (1 - e^{-\gamma}) \cdot e^{-\gamma - 3r_1(1-r_2)\eta\alpha^2} \cdot e^{-\gamma - 3r_1 r_2 \eta\alpha^2}$$

$$P(2|0) = \frac{1}{2}e^{-\gamma} \cdot e^{-\gamma} \cdot (1 - e^{-\gamma}) + e^{-\gamma} \cdot (1 - e^{-\gamma}) \cdot (1 - e^{-\gamma - 3r_1 r_2 \eta\alpha^2}) + (1 - e^{-\gamma}) \cdot e^{-\gamma - 3r_1(1-r_2)\eta\alpha^2} \cdot (1 - e^{-\gamma - 3r_1 r_2 \eta\alpha^2}) + (1 - e^{-\gamma}) \cdot (1 - e^{-\gamma - 3r_1(1-r_2)\eta\alpha^2})$$

$$P(0|1) = e^{-\gamma - 3(1-r_1)\eta\alpha^2} \cdot e^{-\gamma - 3r_1(1-r_2)\eta\alpha^2} \cdot e^{-\gamma - 3r_1 r_2 \eta\alpha^2}$$

$$P(1|1) = \frac{1}{2}e^{-\gamma - 3(1-r_1)\eta\alpha^2} \cdot e^{-\gamma - 3r_1(1-r_2)\eta\alpha^2} \cdot \left(1 - e^{-\gamma - 3r_1 r_2 \eta\alpha^2}\right) + e^{-\gamma - 3(1-r_1)\eta\alpha^2} \cdot (1 - e^{-\gamma - 3r_1(1-r_2)\eta\alpha^2}) \cdot e^{-\gamma} + (1 - e^{-\gamma - 3(1-r_1)\eta\alpha^2}) \cdot e^{-\gamma} \cdot e^{-\gamma}$$

$$P(2|1) = \frac{1}{2}e^{-\gamma - 3(1-r_1)\eta\alpha^2} \cdot e^{-\gamma - 3r_1(1-r_2)\eta\alpha^2} \cdot \left(1 - e^{-\gamma - 3r_1 r_2 \eta\alpha^2}\right) + e^{-\gamma - 3(1-r_1)\eta\alpha^2} \cdot (1 - e^{-\gamma - 3r_1(1-r_2)\eta\alpha^2}) \cdot (1 - e^{-\gamma}) + (1 - e^{-\gamma - 3(1-r_1)\eta\alpha^2}) \cdot (1 - e^{-\gamma})$$

$$P(0|2) = e^{-\gamma - 3(1-r_1)\eta\alpha^2} \cdot e^{-\gamma - 3r_1(1-r_2)\eta\alpha^2} \cdot e^{-\gamma - 3r_1 r_2 \eta\alpha^2}$$

$$P(1|2) = \frac{1}{2}e^{-\gamma - 3(1-r_1)\eta\alpha^2} \cdot e^{-\gamma - 3r_1(1-r_2)\eta\alpha^2} \cdot \left(1 - e^{-\gamma - 3r_1 r_2 \eta\alpha^2}\right) + e^{-\gamma - 3(1-r_1)\eta\alpha^2} \cdot (1 - e^{-\gamma - 3r_1(1-r_2)\eta\alpha^2}) \cdot e^{-\gamma - 3r_1 r_2 \eta\alpha^2} + (1 - e^{-\gamma - 3(1-r_1)\eta\alpha^2}) \cdot e^{-\gamma - 3r_1(1-r_2)\eta\alpha^2} \cdot e^{-\gamma - 3r_1 r_2 \eta\alpha^2}$$

$$P(2|2) = \frac{1}{2}e^{-\gamma - 3(1-r_1)\eta\alpha^2} \cdot e^{-\gamma - 3r_1(1-r_2)\eta\alpha^2} \cdot \left(1 - e^{-\gamma - 3r_1 r_2 \eta\alpha^2}\right) + e^{-\gamma - 3(1-r_1)\eta\alpha^2} \cdot (1 - e^{-\gamma - 3r_1(1-r_2)\eta\alpha^2}) \cdot (1 - e^{-\gamma - 3r_1 r_2 \eta\alpha^2}) + (1 - e^{-\gamma - 3(1-r_1)\eta\alpha^2}) e^{-\gamma - 3(1-r_1)\eta\alpha^2} \cdot (1 - e^{-\gamma - 3r_1 r_2 \eta\alpha^2}) + (1 - e^{-\gamma - 3(1-r_1)\eta\alpha^2}) \cdot (1 - e^{-\gamma - 3(1-r_1)\eta\alpha^2})$$

(a-2)

By using these channel matrixes, prior probabilities are optimized according to the mutual information or cutoff rate.

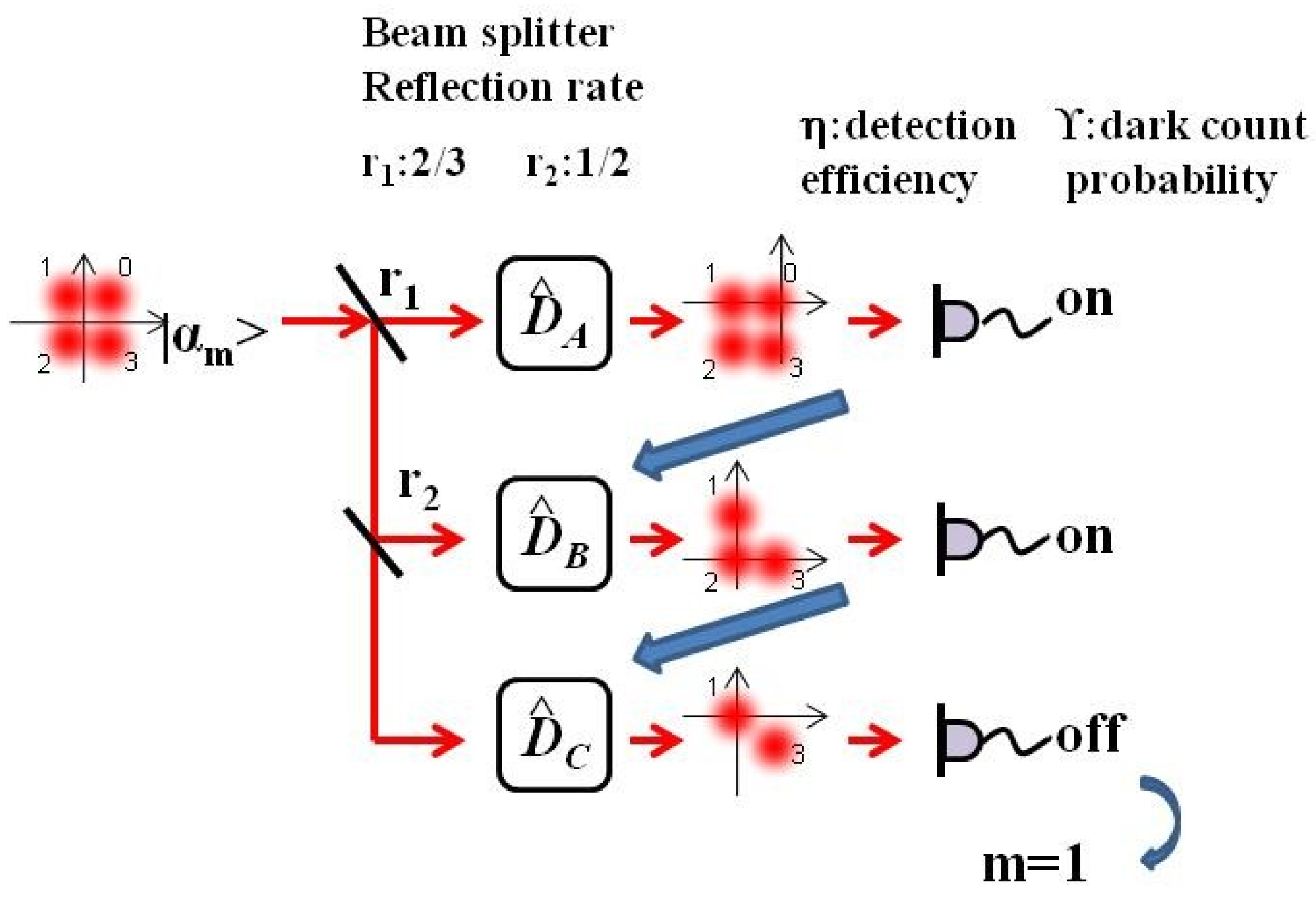

Figure A1. Example of decision process of displacement receiver. Character 'D' in means displacement operator. Displacement value and final decision are determined in accordance with signals of single photon detectors. Here, receiver estimates input signal state, m = 1.